# Improvement of temperature uniformity of induction-heated T-shape susceptor for high-temperature MOVPE

Kuang-Hui Li, Hamad S. Alotaibi, Xiaohang Li*

*King Abdullah University of Science and Technology (KAUST)*

*Advanced Semiconductor Laboratory*

*Thuwal, 23955-6900, Saudi Arabia*

**Corresponding Author: xiaohang.li@kaust.edu.sa*

ABSTRACT

The induction heating is a common method applied in metalorganic vapor phase epitaxy (MOVPE) especially for higher-temperature growth conditions. However, compared to the susceptor heated by the multiple-zone resistant heater, the inductive-heated susceptor could suffer from severe thermal non-uniformity issue. In this simulation study, we propose to employ a T-shape susceptor design with various geometric modifications to significantly improve the substrate temperature uniformity by manipulating thermal transfer. Specifically, the thermal profile can be tailored by horizontal expansion and vertical elongation of the susceptor, or forming a cylindrical hollow structure at the susceptor bottom cylinder. Three optimized designs are shown with different temperature uniformity as well as various induction heating efficiencies. The temperature variation of the entire substrate surface can be less than 5 °C at ~1900 °C with high induction heating efficiency after applying the proposed techniques.

## 1. Introduction

Ultra-wide bandgap III-nitride materials including AlN, BN, and their alloys with other group-III elements are promising for optoelectronics and power electronics applications.[1,2] High material quality of these alloys is essential for device performance and investigation of material properties. However, it has been challenging to realize it especially on commercially-viable foreign substrates including sapphire and silicon primary due to large lattice mismatch. The metalorganic vapor phase epitaxy (MOVPE) is the most common method for growing III-nitride materials. To improve quality of the ultra-wide bandgap III-nitride materials such as AlN, MOVPE growers have employed various methods including precursor pulsing to enhance adatom movement and patterned substrates to leverage lateral coalescence.[3] Another notable method is to apply extremely high temperature (EHT) ( >1600 °C) to enhance adatom mobility and suppress undesirable impurity incorporation, which has led to greatly improved material quality.[4,5]

However, existing commercial MOVPE systems with the resistant heater may not be suitable for long-term and low-cost EHT operation. The filament of the resistant heater is usually made of refractory metals such as tantalum (Ta), tungsten (W), rhenium (Re), or their alloys which can sustain high temperature while emitting thermal radiation. Ideally, the heated susceptor can absorb all the thermal radiation and reach thermal equilibrium temperature according to its emissivity and absorptivity ratio according to the Kirchhoff's law of thermal radiation.[6] However, in reality the susceptor both absorbs and reflects thermal radiation. The susceptor reflects even more thermal radiation at higher temperatures. Therefore, the filament usually needs to be several hundred degrees higher than the susceptor target temperature; otherwise, the susceptor could not reach the target temperature. At higher temperatures, the filament has thermal expansion and can cause warpage and possibly short circuit. Moreover, though refractory metals have high melting point, extreme heating-and-cooling cycles can cause thermal fracture and may break the filament eventually.[7]

Compared to the resistant heater, the induction heater has several advantages due to a different working principle. The induction coil generates alternating magnetic field and the susceptor induces Eddy current accordingly, i.e. inductive coupling. The Eddy current causes the Joule heating effect on the susceptor and heats up the susceptor. Unlike the resistant heater, such energy transfer mechanism is independent of temperature, meaning there is no thermal radiation reflection, warpage, or lifetime issue. That’s the reason for the induction heater to have good heating efficiency and higher reliability than the resistant heater. Nevertheless, the induction heater also has drawbacks. The inductive coupling efficiency between the induction coil and the susceptor is affected by induction coil geometry, susceptor geometry, susceptor material, and frequency of AC power load.[8] Furthermore, the induction heater could result in severe temperature non-uniformity for conventional column-shape susceptors as compared to the resistant heater. Large temperature non-uniformity can cause problems because it affects metalorganic compound pyrolysis efficiency, material composition, growth rate, adatom mobility, and wafer curvature.

The resistant heater users can apply the multi-zone technique[9,10] to tune the substrate temperature uniformity. However, it is difficult to apply the multi-zone technique for the induction heater. There have been studies that propose techniques to improve substrate temperature uniformity of the induction-heated susceptor. But these techniques are often complicated or not applicable for EHT.[11-16] In the previous report, an MOVPE reactor design was proposed by having the induction coil placed around the bottom cylinder under the top plate of a T-shape susceptor as shown in Figure 1(a).[17] Hence, the magnetic field is nearly fully coupled to the susceptor to greatly improve induction heater efficiency at EHT and allow the use of small susceptor-gas inlet distance because of the magnetic shielding effect of the T-shape susceptor.[17] Thus, the proposed reactor could reach higher temperatures and possess lower parasitic reaction rates for Al- and B-containing metalorganic precursors which are desirable for MOVPE processes of the ultra-wide bandgap III-nitride materials. Despite these technical advantages, the T-shape susceptor also suffers from the temperature non-uniformity issue that ought to be addressed for growing high quality and uniform epitaxial wafers.

In this work, the substrate temperature uniformity of the T-shape susceptor has been studied. The substrate temperature uniformity can be improved via controlling the heat transfer path by vertically elongating the susceptor, horizontally expanding the susceptor, or forming a cylindrical hollow structure on the bottom of the susceptor, as shown in Figure 1(b). The induction heating efficiency has also been investigated.

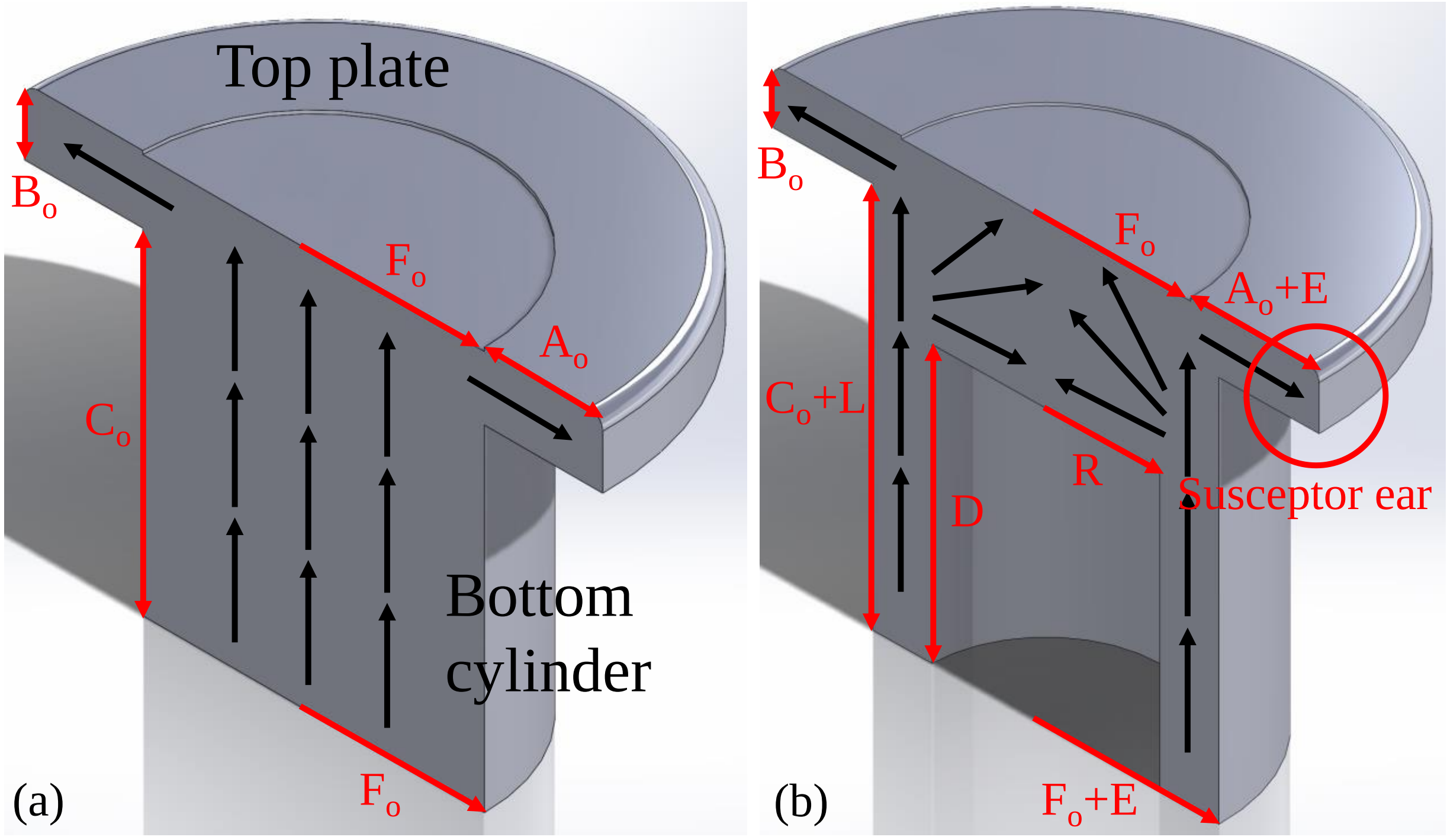

Figure 1 (a) The reference T-shape susceptor with set dimensions of $A_o$ = 1.8 cm, $B_o$ = 1 cm, $C_o$ = 5 cm, and $F_o$ = 2.5 cm (one inch). (b) The optimized T-shape susceptor with geometric variables. Both susceptors can accommodate a two-inch susceptor. Variables R and D in (b) are the radius and depth of the cylindrical hollow structure, respectively. Variable E is the radius increase of the susceptor and variable L is the height increase of the bottom cylinder. The black arrows indicate the heat transfer path.

## 2. Materials and Methods

The heat transfer study was carried out using the cylindrical symmetry due to the geometric of the susceptors. The susceptors were assumed to accommodate a two-inch substrate. Similar works can be conducted on the larger susceptors and thus they are not included in this report. The reference T-shape susceptor in Figure 1(a) is a solid piece comprising a top plate and a bottom cylinder with set dimensions where the bottom cylinder is directly below the two-inch substrate pocket. The optimized T-shape susceptor in Figure 1(b) is similar to the reference T-shape susceptor in terms of the overall shape but have four geometric dimension variables: R and D are radius and height of the cylindrical hollow structure within the bottom cylinder; E is the radial increase of the top plate and the bottom cylinder; and L is the height increase of the bottom cylinder. The position of induction coil can influence the induction coupling. For the reference and optimized T-shape susceptors, the edge of the lowest coil always align with the bottom surface of the bottom cylinder.

The numerical analysis were conducted by using the finite element analysis of the COMSOL Multiphysics 4.3a. There were139,656 triangular elements and 191,043 degrees of freedom included in the mesh. Heat transfer by the induction heating, conduction, and thermal radiation was calculated by the build-in models.[18] In this study, the frequency was fixed at 10 kHz. Most EHT susceptors are made of (crystalline or amorphous) graphite coated by silicon carbide (SiC) or tantalum carbide (TaC). However, in this work, the proposed T-shape susceptor is made of tungsten (W) due to its low cost, high melting point, and superior isotropic electric conductivity and thermal conductivity. TaC is an excellent material for induction-heated high temperature applications but TaC has higher cost than graphite and W. The crystalline graphite has in-plane and out-of-plane lattice planes, which makes it an anisotropic material.[19,20,21] The in-plane electric and thermal conductivities are similar to the tungsten's, depending on the quality of graphite. However, the out-of-plane electric and thermal conductivities are inferior to the tungsten's. Such anisotropic properties affect inductive coupling efficiency and heat transfer. The amorphous graphite is a porous material with poor electric and thermal conductivity. It can lead to poor inductive coupling efficiency that is detrimental for heating efficiency at EHT. Hence, tungsten is a good candidate for inductively-heated high-temperature and low-cost susceptors. *All the physical quantities required in the simulation can be found in the previous report*[17] *and CRC Handbook of Chemistry and Physics.*[22] *For tungsten susceptor, the parameters of resistivity are* $\rho_W = \rho_{W0}[1+\alpha_W(T-T_0)]$, *where* $T_0 = 273$ *K,* $\alpha_W = 5.7\times10^{-3}$ $K^{-1}$, *and* $\rho_{W0} = 4.63\times10^{-8}$ *Ω-m; the parameters of thermal conductivity are* $k_W = \frac{1}{A_W+B_W(T-T_0)}$, *where* $A_W = 6.2\times10^{-3}$ *m-K/W and* $B_W = 3\times10^{-6}$ *m/W. For copper coil, the parameters of resistivity are* $\rho_{Cu} = \rho_{Cu0}[1+\alpha_{Cu}(T-T_0)]$, *where* $T_0 = 273$ *K,* $\alpha_{Cu} = 4.68\times10^{-3}$ $K^{-1}$, *and* $\rho_{Cu0} = 1.52\times10^{-8}$ *Ω-m; the parameters of thermal conductivity are* $k_{Cu} = \frac{1}{A_{Cu}+B_{Cu}(T-T_0)}$, *where* $A_{Cu} = 2.5\times10^{-3}$ *m-K/W and* $B_{Cu} = 5\times10^{-7}$ *m/W. For molybdenum supporter, the parameters of resistivity are* $\rho_{Mo} = \rho_{Mo0}[1+\alpha_{Mo}(T-T_0)]$, *where* $T_0 = 273$ *K,* $\alpha_{Mo} = 5.42\times10^{-3}$ $K^{-1}$, *and* $\rho_{Mo0} = 4.78\times10^{-8}$ *Ω-m; the parameters of thermal conductivity are* $k_{Mo} = \frac{1}{A_{Mo}+B_{Mo}(T-T_0)}$, *where* $A_{Mo} = 7.4\times10^{-3}$ *m-K/W and* $B_{Mo} = 2\times10^{-6}$ *m/W. For Stainless steel showerhead and bottom flange, the parameters of resistivity are* $\rho_{SS} =$

*$\rho_{SS0}[1+\alpha_{SS}(T-T_0)]$, where $T_0$ = 273 K, $\alpha_{SS}$ = 5.84×10$^{-4}$ K$^{-1}$, and $\rho_{SS0}$ = 7.5×10$^{-7}$ Ω-m; the parameters of thermal conductivity are $k_{SS} = \frac{1}{A_{SS}+B_{SS}(T-T_0)+C_{SS}(T-T_0)^2}$, where $A_{SS}$ = 6.8×10$^{-2}$ m-K/W, $B_{SS}$ = −5×10$^{-5}$ m/W, and $C_{SS}$ = 2×10$^{-8}$ m/W-K. For the zirconium oxide thermal insulator, the parameters of thermal conductivity are $k_{ZrO_2} = \frac{1}{A_{ZrO_2}+B_{ZrO_2}(T-T_0)+C_{ZrO_2}(T-T_0)^2}$, where $A_{ZrO2}$ = 1.38×10$^{-1}$ m-K/W, $B_{ZrO2}$ = 2×10$^{-4}$ m/W, and $C_{ZrO2}$ = −6×10$^{-8}$ m/W-K. The emissivity of polished metal is around 0.01 to 0.05 and zirconium oxide thermal insulator is 0.95 at room temperature; however, the emissivity of the metals rises to 0.2 at high temperature. For simplicity, the emissivity of the metals is fixed at 0.2 in the simulation. Sapphire is chosen as the substrate in the simulation. Sapphire has the lowest thermal conductivity (~25 W/m-K) among a few common high-melting-point substrates. High thermal conductivity substrate such as SiC (~360 W/m-K) has better temperature uniformity than sapphire. Sapphire is the worst-case scenario for temperature uniformity. Other substrate won't have temperature uniformity problem if sapphire can achieve temperature uniformity by the techniques in this study. The reactor pressure is kept at 50 Torr close to the ones used to grow AlN in a common MOVPE process today. The gas flow is not considered due to negligible impact on the susceptor temperature. The T-shape susceptor rotation is not included in the simulation since the T-shape susceptor and the induction coil are both cylindrically symmetric. Rotation neither affects the induction coupling efficiency nor changes the substrate temperature profile.*

The temperature distribution on the induction-heated susceptor depends on heat transfer. For the induction heating modeling, the governing equation is:

$$[\nabla^2 + \mu_0\mu_r(\epsilon_0\epsilon_r\omega^2 - i\sigma\omega)]\vec{A} = \mu_0\mu_r(\sigma + i\epsilon_0\epsilon_r\omega)\frac{V_{coil}}{2\pi R}\hat{\phi}, \qquad (1)$$

where $i$ is imaginary number, $\sigma$ is the electrical conductivity, $\omega$ is angular frequency of alternating current, $\eta$ is resistivity of material, $\epsilon_0$ is electrical permeability at free space, $\epsilon_r$ is relative electrical permittivity, $\vec{A}$ is magnetic vector potential, $\mu_0$ is magnetic permeability at free space, and $\mu_r$ is relative magnetic permeability. The induction coil was modeled as torus shape; therefore, the electrical field of the induction coil is $\vec{\nabla}V = \frac{V_{coil}}{2\pi R}\hat{\phi}$, where $R$, $V_{coil}$, and $\hat{\phi}$ are the radius of the induction coil, the electric potential, and the unit vector, respectively.

For thermal conduction modeling, the governing equation is:

$$\rho C_P \frac{\partial T}{\partial t} + \rho C_P \vec{u} \cdot \vec{\nabla} T = \vec{\nabla} \cdot \left(k \vec{\nabla} T\right) + Q, \qquad (2)$$

where $\rho$ is density, $C_P$ is the specific heat capacity at a constant pressure, $T$ is absolute temperature, $t$ is time, $\vec{u}$ is velocity vector, $k$ is thermal conductivity, and $Q = \frac{1}{2}\mathrm{Re}\left(\vec{J} \cdot \vec{E}\right)$ is the power generated by the Eddy current.

For thermal radiation modeling, the governing equations are:

$$E_b(T) = \varepsilon \sigma T^4, \qquad (3)$$

$$(1 - \varepsilon)G = J - E_b(T), \qquad (4)$$

$$-\vec{n} \cdot \vec{q} = G - J, \qquad (5)$$

where $\sigma$ is the Stefan-Boltzmann constant, $E_b(T)$ is the blackbody hemispherical total emissive power, $\varepsilon$ is the emissivity of the material, $G$ is incoming radiative heat flux, $J$ is the total outgoing radiative heat flux, $\vec{n}$ is the normal unit vector on the boundary, and $\vec{q}$ is the radiation heat flux vector.

From the simulation results, the radius and the depth of the hollow structure, and horizontal expansion and vertical elongation of the susceptor can significantly influence the substrate temperature uniformity. The mechanism and optimized parameters for the T-shape susceptor will be discussed thoroughly in the next sections.

## 3. Results and discussion

To quantify the substrate temperature uniformity, we define a value called the *Uniformity Length* (UL) as *the distance from the substrate center to the farthest point within which the substrate surface temperature variation is equal or less than 5 °C*. Thus, the larger the UL, the better the substrate temperature uniformity is. For a two-inch substrate, the maximum UL in the ideal situation is one inch where the temperature difference of the entire substrate is less than 5 °C. Although temperature variation on a wafer in a state-of-the-art MOVPE reactor could be less than 5 °C at lower growth temperatures such as ~1000 °C for InGaN light emitters, the threshold of 5 °C was reasonable because the target is EHT in this study. In other words, a 5 °C difference represented a very small, if not negligible temperature non-uniformity budget at EHT. Figure 2(a) and 2(b) include a convex and a camel-back temperature line profile, respectively, where the UL's are shown as examples. In the following sections, the simulation results related to Figure 3, Figure *5*, and Figure *6* have the same induction heating power of 6.5 kW, but the substrate average temperature is not the same. It is because heat transfer path and induction heating coupling are geometry dependent.

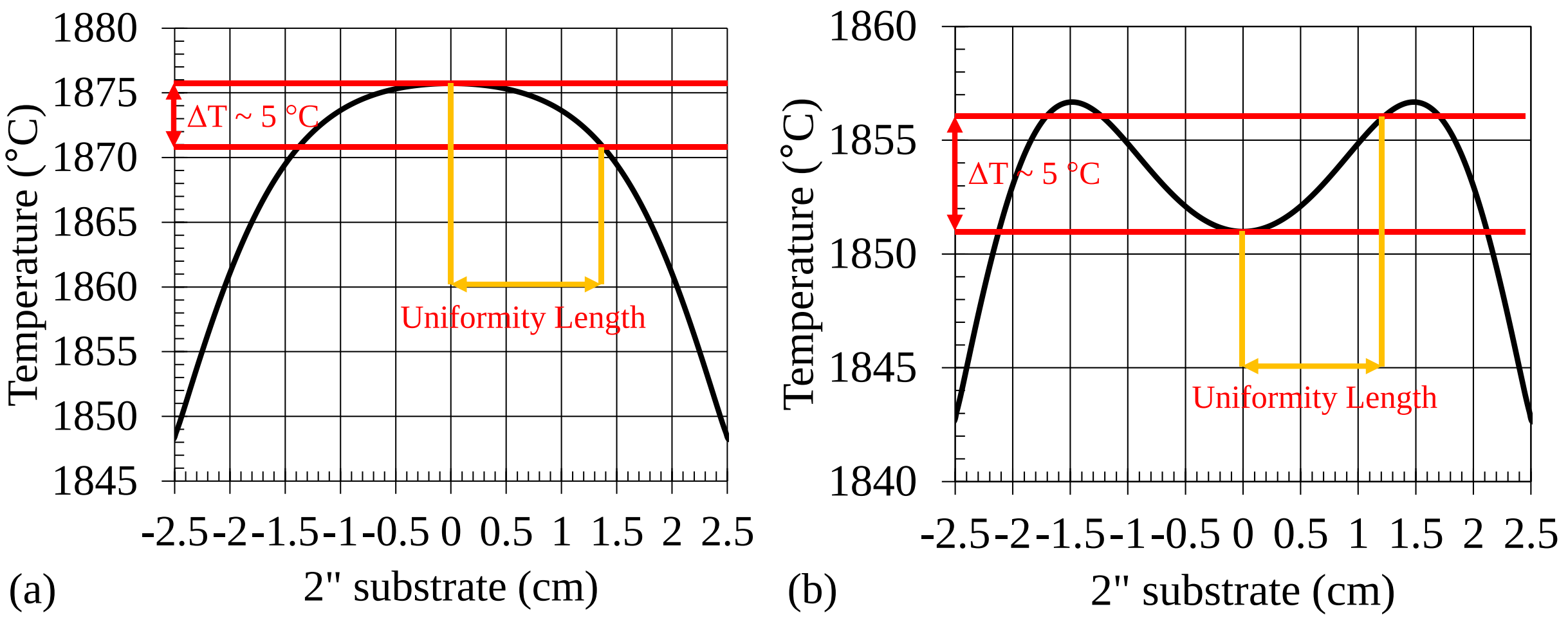

Figure 2 (a) and (b) show the UL's with convex and camel-back substrate surface temperature line profiles, respectively. The zero on the *x*-axis represents the substrate center.

### 3.1 Temperature profile of the reference T-shape susceptor

The cross-sectional temperature profile of the reference T-shape susceptor in the reactor is shown in Figure 3(a). Due to the reactor's axial symmetry, only half of the cross section is shown. The detail reactor configuration can be find elsewhere.[17] The bottom cylinder temperature is higher than the top plate temperature, because the heat transfer is mainly from the bottom cylinder to the top plate. To keep such heat transfer path, the heat transferred downward to the susceptor supporter (made of Molybdenum) and heat released by thermal radiation have to be reduced. Otherwise, these heat sinks will affect the heating efficiency of the susceptor. To reduce the heat sink, several thermal insulators (made of zirconia)[23] are placed on the lateral and bottom sides of the bottom cylinder. The lateral thermal insulator not only reduces thermal radiation, but also protects the induction coil from thermal radiation. The bottom thermal insulator blocks the heat transferring downward to the susceptor supporter.

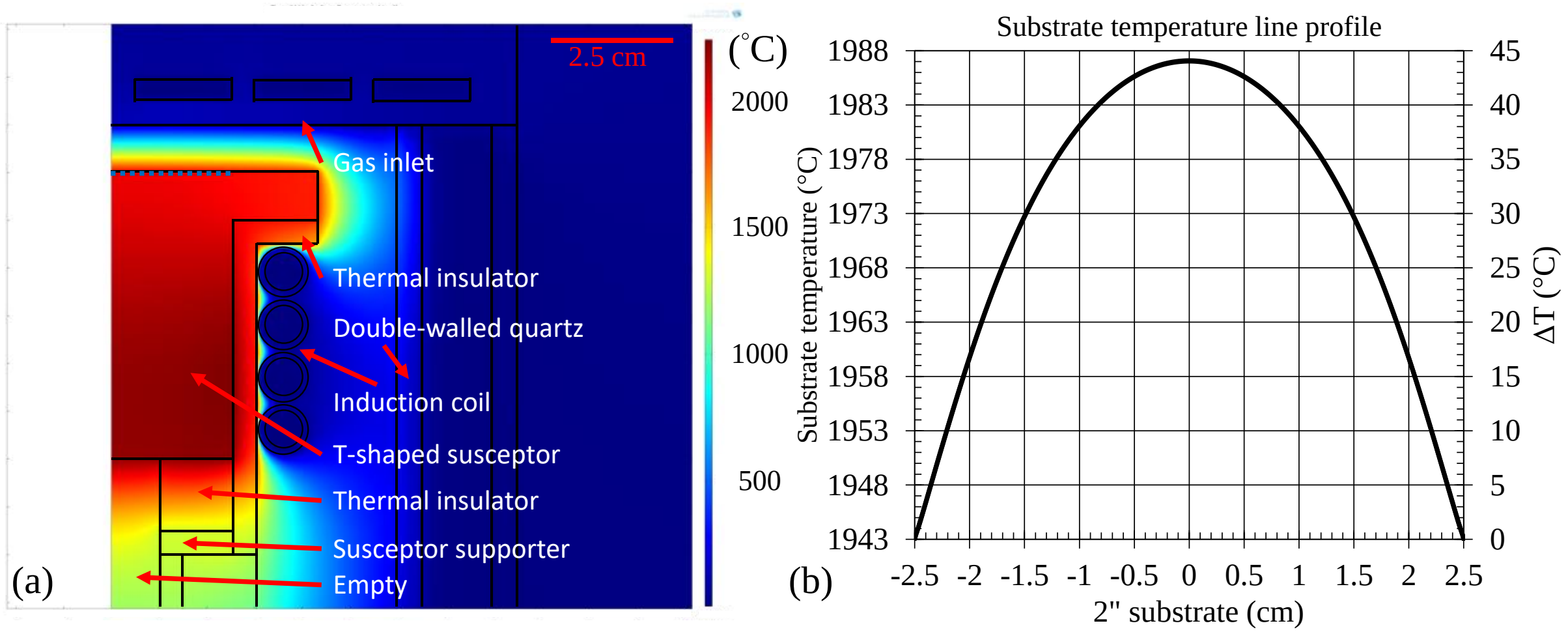

Figure 3 (a) The cross-sectional temperature profile of half of the reference T-shape susceptor in the reactor. (b) The substrate line temperature line profile of the two-inch substrate measured from the blue dash line in (a) with $T_{center}$ of 1987 ˚C.

The substrate temperature line profile is shown in Figure 3(b). The average substrate temperature ($T_{average}$) is 1972 °C with the standard deviation ($\sigma$) of 13.5 °C. The temperature difference ($\Delta T = T_{center} - T_{edge}$) between the susceptor center ($T_{center}$) and the susceptor edge ($T_{edge}$) is as large as 45 °C which is not acceptable. The red curve of Figure 4 shows that Figure 4ΔT a quadratic function of $T_{center}$. When $T_{center}$ is 1000 °C, ΔT is ~5 °C which is still acceptable. However, ΔT rapidly increases to over 25 °C above the EHT, suggested that the reference T-shape susceptor design be modified to be applicable for acceptable uniformity at EHT.

To develop techniques improving the substrate temperature uniformity, understanding the induction heating mechanism and the heat transfer in the T-shape susceptor is important. Based on classical electrodynamics, EM waves only reach a certain depth below a conductor surface and the depth is defined as the skin depth ($\delta$), which can be calculated by the following formula,[24, 25, 26]

$$\delta = \sqrt{\frac{\rho_\eta}{\pi f \mu_o \mu_r}} \sqrt{\sqrt{1 + (2\pi f \rho_\eta \epsilon_0 \epsilon_r)^2} + 2\pi f \rho_\eta \epsilon_0 \epsilon_r}, \quad (6)$$

where $f$ is the frequency of the alternating current, $\epsilon_0$ is the electrical permeability at free space, $\epsilon_r$ is the relative electrical permittivity, $\mu_0$ is the magnetic permeability at free space, $\mu_r$ is the relative magnetic permeability, and $\rho_\eta$ is the resistivity of the conductor at temperature $\eta$. Because $2\pi f \rho_\eta \epsilon_0 \epsilon_r$ is a small quantity for common induction heaters frequency and metal, Equation (6) can be further reduced to:

$$\delta(\eta) \approx \sqrt{\frac{\rho_\eta}{\pi f \mu_0 \mu_r}} \approx 503.29\sqrt{\frac{\rho_0[1+\alpha(\eta-\eta_0)]}{f}}, \quad (7)$$

where $\rho_0$ is the reference resistivity of the conductor at temperature $\eta_0$ (300K) and $\alpha$ is the temperature coefficient of resistivity. Tungsten is a paramagnetic material which can be consider as low-level magnetization, and paramagnetic property can be described by the Curie's Law. Therefore without large deviation, $\mu_r$ can be approximately considered as one. From Equation (7), $\delta$ depends on the electrical properties of the conductor, the frequency of the induction heater, and the temperature of the susceptor.

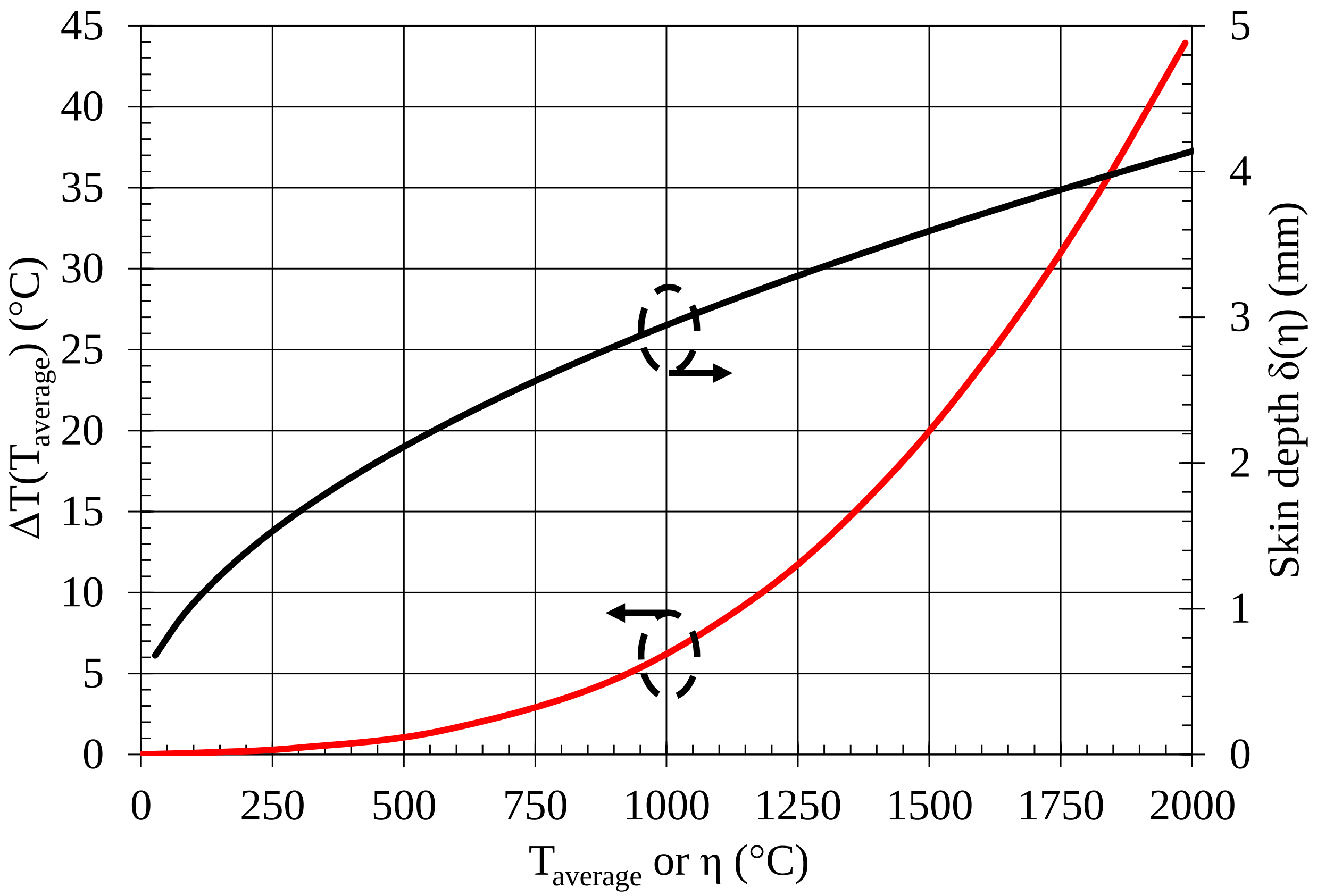

Figure 4 (Red curve) The temperature difference (ΔT) between $T_{center}$ and $T_{edge}$ of the reference T-shape susceptor as a function of $T_{average}$. (Black curve) Skin depth as a function of the temperature $\eta$.

The skin depth vs temperature $\eta$ was further calculated and shown in Figure 4. The skin depth gradually increases from 1.1 to 4 mm when temperature increases from 100 to 2000 °C. The magnetic field intensity decays exponentially when the magnetic field penetrates into the T-shape susceptor ($\sim e^{x/\delta}$).[25,26] When the magnetic field penetrates one skin depth distance from the surface, the intensity decays to 36.7% ($\sim e^{-1}$), and decays to 13.5% ($\sim e^{-2}$) and 4.9% ($\sim e^{-3}$) when the penetration distances are two and three skin depth, respectively. Therefore within the distance of three skin depths from the surface, the bottom cylinder will induce most of the Eddy current and generate an internal magnetic field against the external magnetic field by the Faraday-Lenz law of induction.[27] The Eddy current encircles the bottom cylinder and generates heat by the Joule-Lenz law. The skin depth at 1900 °C is 4 mm from Figure 4, which means that from the bottom cylinder surface to 1.2 cm below, the Eddy current will be induced to generate heat. Once the generated heat transfers to the top plate surface, the temperature of the outer region of the bottom cylinder near the surface is lower than that of the inner region. It is because the outer region will release heat by radiating thermal radiation and conducting to the thermal insulator. Furthermore, when the heat approaches the top plate surface, part of the heat goes to the susceptor ear (Figure 1), making the outer region of the bottom cylinder release more heat.

The heat transfer behaviors explain that the two-inch substrate has higher $T_{center}$ and lower $T_{edge}$ [Figure 3(b)], and the temperature difference between the center and the edge increases as the average temperature goes higher. The substrate temperature line profile [Figure 3(b)] has a UL of 0.92 cm, which corresponding to 13.5% temperature uniformity on the substrate surface [ $\left(\frac{0.92\ cm}{2.5\ cm}\right)^2 \sim 13.5\%$ ]. Such uniformity is not acceptable. However, it can be improved by geometric modification techniques in Section 3.2.

### 3.2 Impacts of geometric options on substrate temperature uniformity

In this section, out of the geometric options, i.e. the formation of the hollow structure (R and D), the radius increases of the top plate and the bottom cylinder (E), and the bottom cylinder elongation (L), only one is implemented at one time while others are the same as the reference susceptor. After the impact of each option is known, it helps further improving the substrate temperature uniformity when multiple variables are involved (Section 3.3). Figure 5(b) presents the substrate temperature line profile at different R and D values, while keeps E and L zero. Figure 5(c) exhibits the substrate temperature line profile evolution at various E values, while R, D, and L are zero. Figure 5(d) shows the impact of L while R, D, and E are zero. The three options are found to significantly impact the heat transfer path and the substrate temperature uniformity.

In Figure 5(b), the substrate temperature line profile shifts downward when R and D increase. Meanwhile, the substrate temperature uniformity gradually improves. For instance, when R is 2 cm and D is 5 cm (blue curve) shows superior substrate temperature uniformity to red and green curves. The explanation is that when there is a hollow structure in the bottom cylinder, the heat transfer is not simply from the entire bottom cylinder to the top plate, since the inner region of the bottom cylinder is empty, as shown in Figure 1(b). When R and D increase, more heat starts to transfer from the side of the bottom cylinder to the center of the bottom cylinder. Such a shift of the heat transfer results in the substrate temperature drop at the center. Since the substrate temperature at the center was higher than the substrate temperature on the edge, the substrate temperature uniformity can be improved by forming the hollow structure.

In Figure 5(c), the substrate temperature line profile evolves with different values of E. The substrate temperature line profile shifts downward with increasing E, while the temperature difference between the center and the edge reduces and achieves acceptable substrate temperature uniformity when E is 1.5 cm. In Section 3.1, the substrate temperature difference between the substrate center and edge can be explained by different heat transfer paths between the bottom cylinder inner and outer regions. Here, a larger diameter keeps the bottom cylinder inner region away from the outer region. Hence, the heat transferring to the top plate surface is mainly from the bottom cylinder inner region. Since the temperature in the bottom cylinder inner region is much uniform than that in the bottom cylinder outer region, the temperature uniformity is improved.

In Figure 5(d), interestingly, the substrate temperature line profile shifts downward without alternating its shape. Also, the profile drops almost linearly (about 30 °C for every 0.5 cm increase in L). This is because the vertical elongation does not change the heat transfer path. The bottom cylinder outer region still has a faster heat lost rate than that of the bottom cylinder inner region. Furthermore, since the induction coil also shifts downward amid the vertical elongation, the heat generated in the bottom cylinder is even farther away from the top plate surface, which makes the heat transfer path longer. Therefore, the substrate temperature line profile shifts downward without improvement or deterioration of the substrate temperature uniformity. By studying how the temperature line profile is affected by R, D, E, and L, it is obvious that R, D, and E can improve the uniformity greatly if appropriate R, D, and E are chosen. On the other hand, L is not useful.

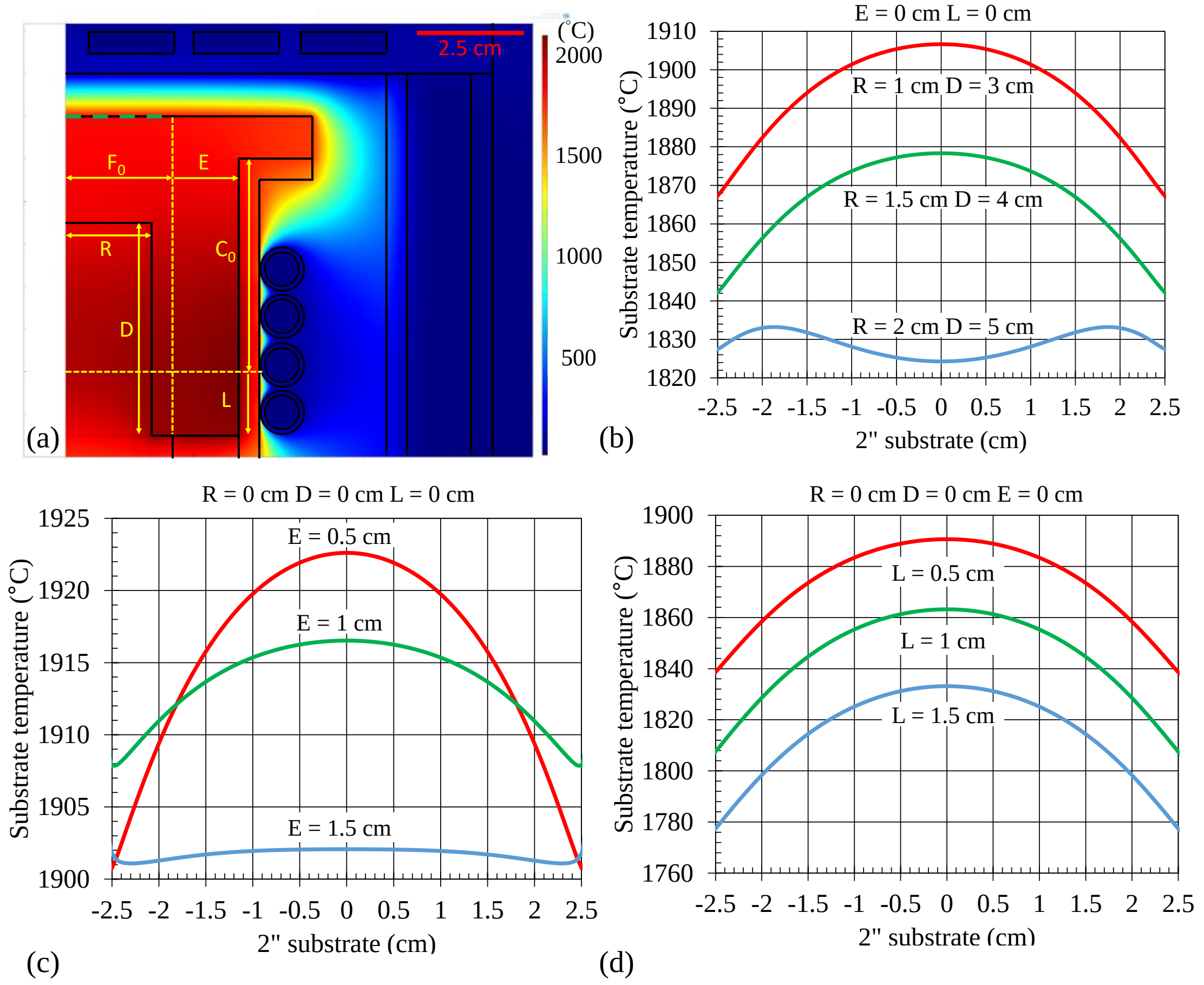

Figure 5 (a) The cross-sectional temperature profile of half of T-shape susceptor in the reactor. The vertical yellow dash line shows the original radius of the bottom cylinder. The horizontal yellow dash line shows the original bottom edge of the bottom cylinder before elongation. The red solid line on the up-right corner is a scale bar. (b), (c), and (d) show different substrate temperature line profiles by adjusting R and D, E, and L, respectively, while fixing the other variables as shown on the top of each figure.

The current section (3.2) discusses the impacts of the three geometric options. But they are not fully optimized even though some examples in Figure 5 show better substrate temperature uniformity. To further improve the substrate temperature uniformity, R, D, and E have to be optimized according to the temperature standard deviation, the UL, and the susceptor volume. In Section 3.3, three optimized designs (Designs 1-3) are shown and discussed with their own benefits and drawbacks.

### 3.3 Comparison between three optimized designs

The optimization follows one major rule: it has to keep the UL as large as possible. The maximum is 2.5 cm. On top of that, if there are multiple optimized setups which leads to the same UL, the susceptor volume is preferably smaller, which is related to heating efficiency at the EHT. It is important to note that the difference in the required induction power may not be large in this study between the two-inch substrate susceptors. But it would be expectedly significant for larger susceptors particularly the ones used for production. The substrate temperature line profile of Design 1 is shown in Figure 6(b). It is apparent that the substrate temperature uniformity matches the criteria ($\Delta T \leq 5$ °C). Its substrate average temperature is 1907 °C, the temperature standard deviation is 1.0 °C, and the UL is 2.5 cm covering the entire substrate. For Design 2 [Figure 6(d)], its substrate temperature uniformity is not as good as Design 1. Design 2 has average substrate temperature of 1835 °C, the temperature standard deviation of 3.1 °C, and the UL of 2.1 cm covering 71% of the substrate surface. The reduced UL is caused by an unfavorable substrate temperature drop (~ 13 °C) near the substrate edge. For Design 3 [Figure 6(e)], it has the substrate average temperature of 1837 °C, the temperature standard deviation of 1.4 °C, and the UL of 2.5 cm covering the entire substrate.

Through optimizing horizontal expansion only, Design 1 seems to be better than Design 2 optimizing the hollow structure only. However, there is a drawback. The substrate temperature uniformity is improved by increasing the radius versus the reference substrate. This causes the volume of Design 1 is roughly twice larger than that of the reference substrate. Thus, Design 1 requires more induction power. For instance, Design 1 requires 6.4 kW at 1750 °C while the reference susceptor needs only 4.4 kW [Figure 7(a)]. The uniformity of Design 2 is poorer. However, it has the closest heating efficiency to that of the reference substrate. For Design 1 and Design 2, there is a tradeoff between the temperature uniformity and the heating efficiency. It is possible to have a compromised design, which means excellent temperature uniformity and high heating efficiency, i.e. Design 3. Design 3 leads to the same UL as Design 1. Since Design 3 has larger standard deviation than Design 1's, Design 1's temperature uniformity is better. But Design 3 has considerably higher heating efficiency than that of Design 1, as shown in Figure 7(b).

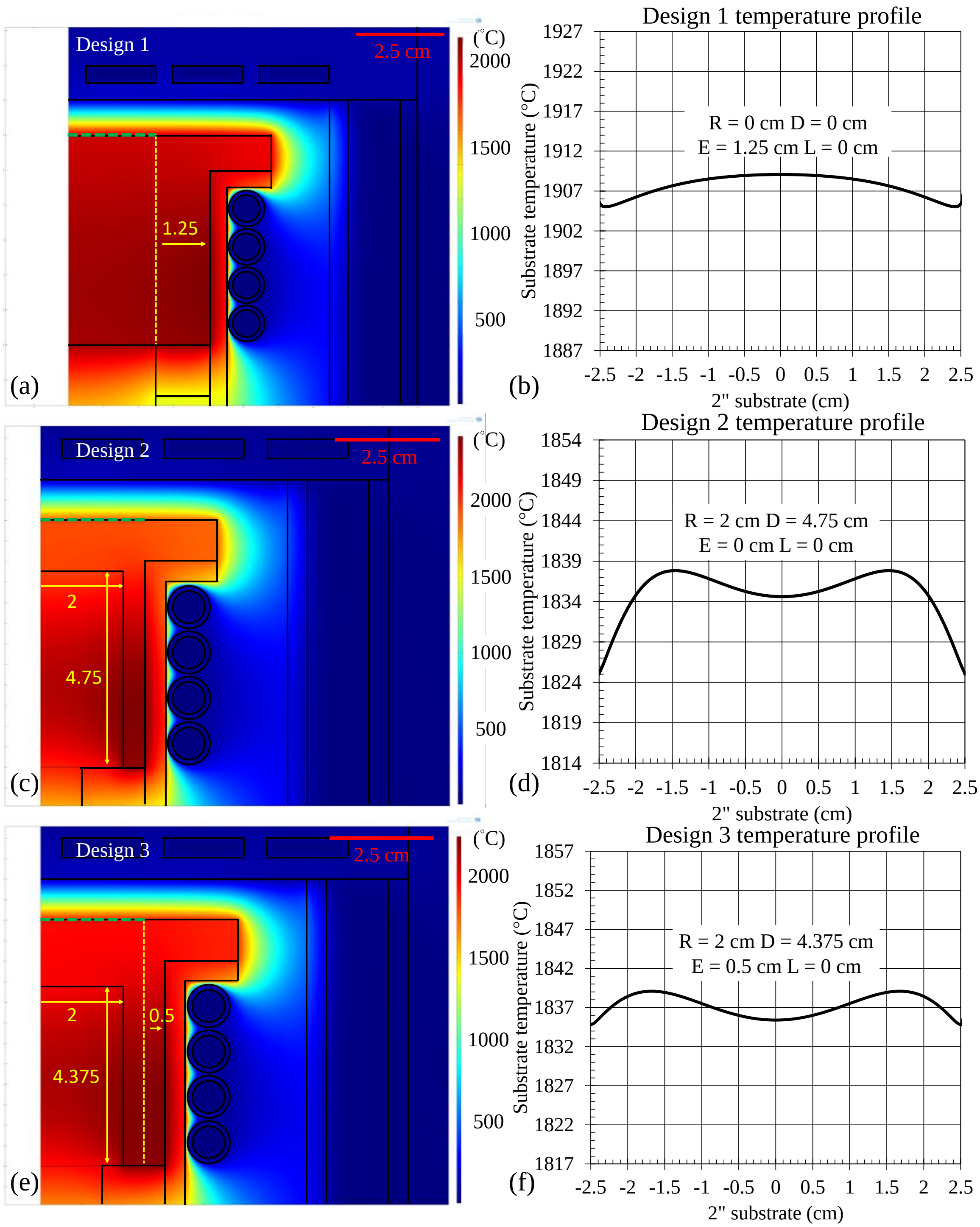

Figure 6 (a), (c), and (e) are the temperature profiles of the optimized T-shape susceptors Design 1–3, respectively. The unit of the numbers is cm. (b), (d), and (f) are the substrate temperature line profiles of Design 1–3, respectively.

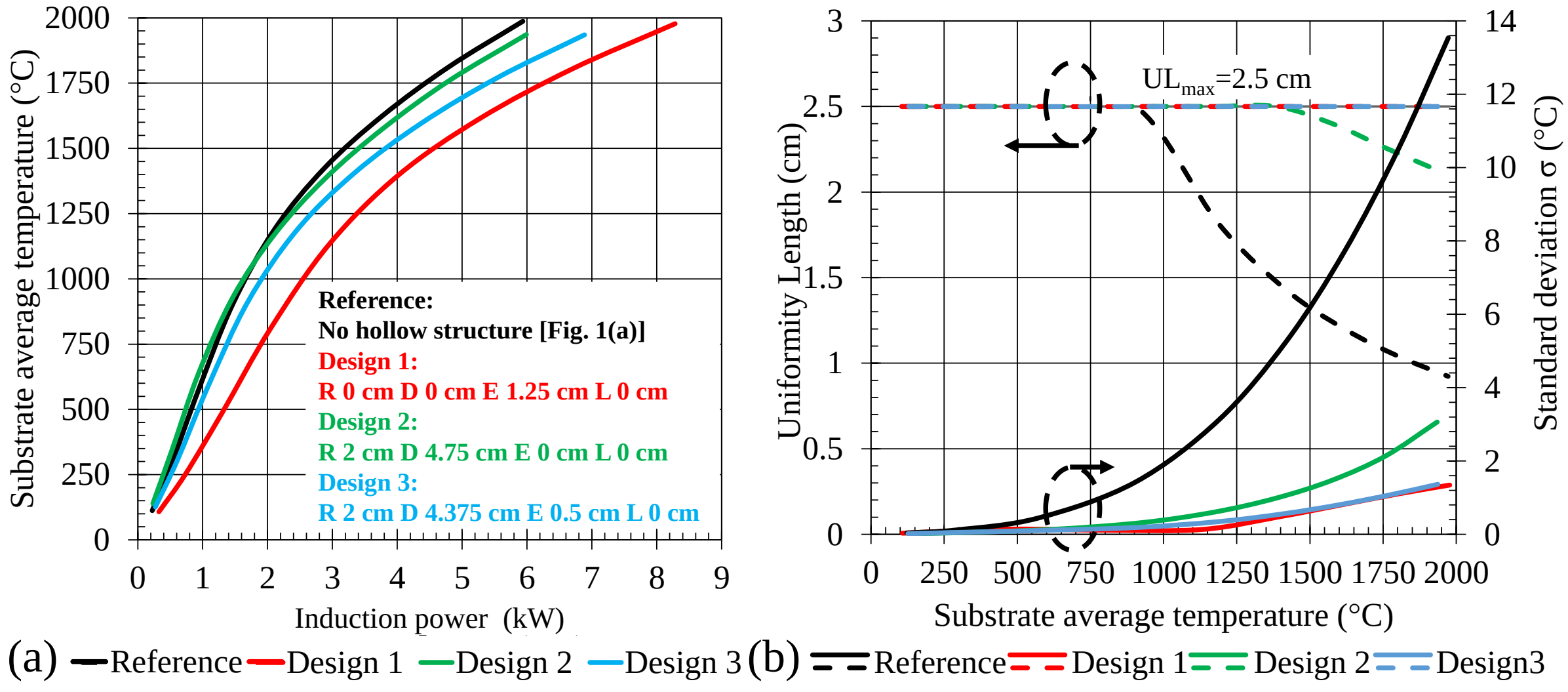

*Figure 7 (a) The substrate average temperature of different designs as a function of the induction power, indicating various heating efficiencies. (b) The UL and the standard deviation of the substrate temperature as a function of the substrate average temperature.*

The UL and the standard deviation of the substrate temperature as a function of the substrate average temperature of the three designs are shown in Figure 7(a). For Design 1 and Design 3, the UL is 2.5 cm amid the entire temperature range indicating their excellent candidacy for the temperature uniformity. For Design 2, the UL is 2.5 cm until reaching temperatures over ~1500 °C, which means that it is perfect for lower temperatures but not good for the EHT. The UL of the reference susceptor starts to decrease at 900 °C which is even lower than the conventional growth temperature of GaN (~1000 °C). The standard deviation increases quadratically versus the average temperature and largely reflects the same phenomena as the UL does. There is a correlation between the UL and the temperature standard deviation: once the substrate temperature standard deviation goes beyond ~1.6 °C, the UL starts to drop. The explanation is that if taking substrate temperature line profile as a Laplace-Gauss distribution, 3σ covers 99.7% of the data points. To match acceptable substrate temperature uniformity, 3σ should be equal to or less than 5 °C ($3\sigma \leq 5$ °C), which gives the result of $\sigma \leq 1.67$ °C.

## 4 Conclusion

In summary, the T-shape susceptor is a candidate for high temperature MOVPE processes but can suffer severe temperature non-uniformity issues. In this study, it is found that the modifications of the susceptor geometric can significantly impact the temperature profile and improve uniformity. Specifically, the radius increase of the susceptor and the formation of the hollow structure of the susceptor bottom cylinder can greatly improve temperature uniformity through manipulating the thermal transfer, while the length increase of the susceptor bottom cylinder can only shift the temperature profile. The geometric modification also causes change in the induction heating efficiency. With the proposed techniques, the T-shape susceptor can exhibit excellent temperature uniformity with temperature variation less than 5 °C at ~1900 °C and high induction heating efficiency.

## 5 Acknowledgement

The authors would like to acknowledge the support of KAUST Equipment Fund BAS/1/1664-01-08, KAUST Baseline BAS/1/1664-01-01, Competitive Research Grant URF/1/3437-01-01, and GCC Research Council REP/1/3189-01-01. In addition, we thank Dr. Gary Tompa from Structured Materials Industries (SMI) for fruitful discussion of the substrate temperature uniformity and the MOVPE designs with induction heating and resistant heating.

## Reference

[1] A. Khan, K. Balakrishnan, and T. Katona, Nat. Photon. 2, 77(2008).

[2] K. Watanabe, T. Taniguchi, T. Niiyama, K. Miya, and M. Taniguchi, Nat. Photon. 3, 591 (2009).

[3] J. Zhang, H. Wang, W. Sun, V. Adivarahan, S. Wu, A. Chitnis, C. Chen, M. Shatalov, E. Kuokstis, J. Yang, A. Khan, J. Electron. Mater. 32, 364 (2003).

[4] A. Rice, A. Andrew, C. Mary, B. Thomas, O. Taisuke, S. Catalin, F. Jeffrey, and S. Michael, J. Cryst. Growth 485, 90 (2018).

[5] N. Fujimoto, T. Kitano, G. Narita, N. Okada, K. Balakrishnan, M. Iwaya, S. Kamiyama, H. Amano, I. Akasaki, K. Shimono, T. Noro, T. Takagi, and A. Bandoh, Phys. Status Solidi C 3, 1617 (2006).

[6] G. Kirchhoff, Monatsberichte der Akademie der Wissenschaften zu Berlin, 1859, 783 (1860).

[7] C. Li, D. Zhu, X. Li, B. Wang, J. Chen, Nucl. Mater. and Ener. 13, 68 (2017).

[8] S. Hu, Q. Wu, J. Li, H. Cao, Y. Zhang, and Z. Li, Mater. Sci. Eng. 322, 022006 (2018).

[9] Y. Qu, B. Wang, S. Hu, X. Wu, Z. Li, Z. Tang, J. Li, and Y. Hu, J. Cent. South Univ. 21, 3518 (2014).

[10] M. Tsai, C. Fang, and L. Lee, Chem. Eng. Process. 81, 48 (2014).

[11] Z. Li, Y. Hao, J. Zhang, L. Yang, S. Xu, Y. Chang, Z. Bi, X. Zhou, and J. Ni, J. Cryst. Growth 311, 4679 (2009).

[12] Z. Li, H. Li, J. Zhang, J. Li, H. Jiang, X. Fu, Y. Han, Y. Xia, Y. Huang, J. Yin, L. Zhang, and S. Hu, Appl. Therm. Eng. 67, 423 (2014).

[13] Z. Li, J. Zhang, J. Li, H. Jiang, X. Fu, Y. Han, Y. Xia, Y. Huang, J. Yin, L. Zhang, and Y. Hao, J. Cryst. Growth 402, 175 (2014).

[14] C. Kim, J. Hong, J. Shim, Y. Won, Y. Kwon, IEEE 11th International Conference on EuroSimE, pp. 1-7 (2010).

[15] Z. Li, H. Li, X. Gan, H. Jiang, J. Li, X. Fu, Y. Han, Y. Xia, J. Yin, Y. Huang, and S. Hu, J. Semicond. 35, 092003 (2014).

[16] Z. Li, H. Li, J. Zhang, J. Li, H. Jiang, X. Fu, Y. Han, Y. Xia, Y. Huang, J. Yin, L. Zhang, S. Hu, Int. J. Heat Mass Transfer 75, 410 (2014).

[17] K.-H. Li, H. Alotaibi, H. Sun, R. Lin, W. Guo, C. Torres-Castanedo, K. Liu, S. Valdes-Galán, and X. Li, J. Cryst. Growth 488, 16 (2018).

[18] COMSOL Multiphysics® User’s Guide VERSION 4.3

[19] P. Wallace, Phys. Rev. 71.9, 622 (1947).

[20] L. Wang, Z. Tamainot-Telto, S. Metcalf, R. Critoph, R. Wang, Appl. Therm. Eng. 30, 1805 (2010).

[21] G. Slack, Phys. Rev. 127, 694 (1962).

[22] W. Haynes, ed. CRC handbook of chemistry and physics. CRC press, 2014.

[23] L. Hu, C. Wang, and Y. Huang., J. Mater. Sci. 45, 3242 (2010).

[24] V. Rudnev, D. Loveless, R. L. Cook, and M. Black (2002). Handbook of Induction Heating. CRC Press.

[25] D. J. Griffiths (1999). Reed College, Introduction to Electrodynamics, Vol. 3.

[26] D. J. John (1999). Classical Electrodynamics.

[27] E. Lenz, Annalen der Physik und Chemie 107, 483 (1843).